\documentclass{cpbtex}

\begin{document}

\begin{CJK*}{GBK}{song}

\title{Modeling of the energy transfer process in microtubules}  

\author{S. Eh. Shirmovsky}
\thanks{ E-mail:~shirmovskiy.sye@dvfu.ru}\\
{Far Eastern Federal University, Institute of Mathematics and Computer Technologies,  Department of Information Security. 10 Ajax settlement, Russkiy Island, Vladivostok, Primorsky Region, Russia 690922. }\\

\date{\today}
\maketitle

\begin{abstract}


In the work the process of the energy transfer in the cell microtubules  is simulated.  A system of tryptophans connected by Coulomb dipole-dipole interaction is discussed as the energy carriers.  The work models the conditions under which the migration of energy along the tryptophan chain in the microtubule is possible.
It was shown  the transfer mechanism has a mixed nature. Thus, within two or three tryptophans with a strong dipole-dipole interaction, the energy transformation process occurs due to an exciton coherent mechanism. In cases of weaker dipole-dipole interaction, the energy transformation process occurs due to an inductive-resonant mechanism. The results of the work allow us  talk about a possible signal function of a microtubules, transmitting signals about processes locally induced in its dipole-dipole structure.  The energy transfer in microtubules has been determined as a quantum phenomenon. 

\end{abstract}
\textbf{Keywords:} microtubule;  quantum biology; nonlinear dynamics; energy transfer\\
\textbf{PACS:}~03.65.-w; 3.65.Yz; 05.45.-a; 87.15.A-

\section{Introduction}
 
The energy transfer in biostructures can be associated with three processes: electron transfer, tunneling effect, and migration of the electron excitation energy, which is not accompanied by electron separation from the donor molecule.  In the later case, for us, inductive-resonant and exciton transfer mechanisms are of great importance.

The transfer of the electron excitation energy from the excited donor molecule to the acceptor molecule is determined by the their interaction features.  In the inductive-resonant mechanism case, the energy transfer occurs due to the Coulomb (dipole-dipole) interaction between the donor and acceptor molecules. The excited electron of the donor molecule interacts with the electron of the acceptor molecule, which are at a lower energy level. In this case, after a short period of time, the electron in the donor returns to the  lower energy level, and the acceptor molecule is excited. As a rule, an inductive-resonant energy transfer mechanism is possible between the singlet levels of the donor and acceptor. However, transitions between the triplet (singlet) levels of the donor and the singlet (triplet) levels of the acceptor are possible too~\cite{rub}~--~\cite{turo}.

The exciton mechanism of energy migration is carried out at higher interaction energies. In this case, the excitation passes to the neighboring acceptor before the transition  to the lower energy  level in the donor occurs. The excitation does not have time to be localized on each molecule separately.  The strong interaction of neighboring molecules causes a correlation of their charge displacements, i.e. the formation of a wave packet covering several molecules at once -- a coherent exciton.

In the work  the mechanisms of energy transfer in the microtubules are investigated.

The main mechanical framework of eukaryotic cells is the cytoskeleton. The cytoskeleton is a set of filamentous protein structures - actin filaments, intermediate filaments and tubulin microtubules. Microtubules are hollow cylindrical tubes with an outer diameter of about 25 nm and an inner diameter of about 14 nm. They consist of tubulin molecules, each of which is a dimer formed by two tightly bound globular subunits.  These subunits are related proteins called $\alpha$- and $\beta$-tubulins {\bf Fig.~1}~\cite{Albetrs}. 

Currently, the study of the mechanical and biological properties of microtubules is relevant. These studies allow us to better understand their significance for the functioning of biosystems. In addition, it also seems relevant to conduct the research of the role of quantum effects. The influence of quantum effects on the functioning of biological systems is currently being studied very extensively. At the moment, their importance in such processes as photosynthesis, charge transfer in DNA has been established~\cite{ham}~--~\cite{42}. The microtubule is also considered as an object for quantum mechanical research~\cite{ham},~\cite{adpe},~\cite{SS21}. 

At the moment, it is a reasonable assumption that quantum  effects in biosystems can be associated with organized quantum processes in $\pi$-electron clouds inside biomolecules~\cite{ham}.  The protein subunits of tubulin that make up the microtubule have a special arrangement of chromophores -- aromatic amino acids that include tryptophans. Due to the high concentration of $\pi$-electrons in tryptophan and the high polarizability of the connected aromatic rings, the amino acid tryptophan is a suitable molecule for energy transfer in microtubules.

The results of three-dimensional crystallographic modeling of the tubulin structure show that there are eight tryptophans in the tubulin molecule -- four in the alpha monomer and four in the beta monomer~\cite{nog}.  Two tryptophans (one from each monomer) are located at the base of the dimer. The tryptophans  are close to the inner surface and are arranged vertically~\cite{hamer}. The interval between the nearest tryptophans is from 11.4  ${\buildrel _{\circ} \over {\mathrm{A}}}$ to 41.6 ${\buildrel _{\circ} \over {\mathrm{A}}}$~\cite{{get}}. If we connect the nearest tryptophans, we can see that the system of tryptophans in a single protofilament creates a path for energy transfer between tryptophans along the protofilament.

At the moment, it is assumed that such structures can be the basis of the microtubule signaling system and perform the functions of information  transmitters. Moreover, it is assumed that the mechanism of information transmission can be exclusively quantum in nature ~\cite{ham}, ~\cite{SS21}, ~\cite{hamer}. 

The crystal-like structure of microtubules makes them attractive candidates for the role of a medium in which quantum excitations are transmitted. It can be assumed that tryptophans located in microtubules can support the process of the energy migration in them.

In the work, the chain of twenty-four tryptophans located in four tabulins was selected as the object for investigation {\bf Fig.~2}.

The work is organized as follows: in Section 2, the model is defined;  in Section 3, the  initial conditions and model parameters are determined; in Section 4, the mechanisms of the energy migration in microtubules are investigated. In the conclusion, the results of the work are determined.  Peculiarities of the energy transformation investigation in a triptophane system are discussed.

\section{Model}

Due to the shift of electron density in tryptophan molecules, they are electric dipoles interacting with each other. The absorption of electromagnetic energy by a tryptophan molecule in a tabulin leads to the transition of tryptophan into an excited state.
Interaction of excited tryptophan molecules with the environment, gives an irreversible character of the energy dynamics in the microtubule. It is known that in quantum mechanics it is impossible to give an adequate description of irreversible processes based on the Schrodinger or Liouville equations. The use of these equations will be valid for describing of the reversible processes occurring in closed quantum systems. Another approaches are used for description of the open quantum systems, interacting with the environment that irreversibly evolve over time - the Born-Markov ~\cite{SS21} and Lindblad~\cite{lbl} equations, Schrodinger equation with dissipative part  -- quantum-classical formalism developed for example in the works~\cite{lakh}~--~\cite{42}, subdynamics theory~\cite{sb1}~--~\cite{sb3}.

Quantum-classical formalism,  has proven itself well for example in describing the migration of charge along the chains of nitrogen bases of DNA. So, in works~\cite{41} DNA consists of two polynucleotide strands was taken into consideration. It was shown that the transfer mechanisms depend on the sequence type and can be either of hopping nature or of superexchange one. In the work~\cite{40} the fact that a DNA molecule is formed by a furanose ring as its sugar, phosphate group and bases was taken into consideration. Based on the model, results were obtained for the probability of a hole location on the DNA sugar-phosphate groups and base sequences. In addition, the quantum-classical formalism has proven itself well in describing  the primary processes of charge transfer in the photoreaction center of plants~\cite{dnaex3}. 

According to this approach, microtubule tubulin tryptophans can be represented as a classical system of harmonic oscillators. The Hamiltonian $H_{Cl}$ of such system  can be represented as the sum of the kinetic term and the interaction term~\cite{lakh}.
\begin{equation}
H_{Cl}=\sum_n M\frac{ \dot{X}_{n}^{\,2}}{2} + \sum_n k\frac{ X_{n}^{\,2}}{2},
\label{hcl}
\end{equation}
where $M$ is the mass of tryptophan; $k$ is elastic constant; $X_n$ is the displacement of $n$-tryptophan from the equilibrium position;  point "$\cdot$" defines the time derivative. Summation being performed over all the tryptophans under consideration.

It is assumed that the tryptophan system can oscillate both due to the movement inside the tabulins and due to fluctuations of the tabulins in the microtubule. Thus, the tryptophan system can be represented as a macro structure whose dynamics is described by the canonical Hamilton's equations.

The excitation in tryptophans in  the work is associated with the dynamics of an electron, therefore it can be considered as a quasiparticle whose dynamics has a quantum nature and obeys the Schrodinger equation.
\begin{align}\label{shr}
i \hbar \frac{\partial | \psi \rangle}{\partial t} = {\hat H}_Q | \psi \rangle,
\end{align}
with $| \psi \rangle$ being a quasiparticle wave function. 

In equation~\eqref{shr} the simulation of excitation transfer along the tryptophan chain is based on the Hamiltonian (operator) of the form~\cite{lakh},~\cite{dnaex3},~\cite{41},~\cite{ent}:
\begin{equation}
\hat H_Q= \sum\limits_n {\omega_{n} | n \rangle \langle n|}+
										\sum\limits_{n \ne n'}	{
											\nu_{n,n'} | n \rangle \langle n'|}, 
\label{qua}
\end{equation}
where $\omega_{n}$ is the excitation energy of $n$-tryptophan; the state $|n\rangle$ denotes the excitation located on $n$ tryptophan; $\nu_{n,n'}$ is the energy of the dipole-dipole interaction between tryptophans. 

The work assumes that $\omega_{n}$ is a variable that can be determined by  the interaction of $n$ tryptophan  with the environment. In other words it is assumed that the energy $\omega_{n}$  is a linear
function of the displacement of the tryptophan from equilibrium positions. 
It should be emphasized that in contrast to work~\cite{SS21}, where the dependence of energy  on the environment is  of static nature, in this case  the excitation energy of tryptophans  depends on the dynamics of tryptophans in the microtubule.
With this in mind, in the work, the excitation energy is represented as a sum~\cite{dnaex3},~\cite{41}
\begin{equation}
\omega_{n}=\omega_{n}^{0}+\alpha_{n}^{'}X_n.
\label{exe}
\end{equation}
where the first term $\omega_{n}^{0}$ is static and corresponds to the excitation energy of the separate, isolated tryptophan. The second term is dynamic, where $\alpha_{n}^{'}$ is the constant of the relationship between the excitation energy of tryptophan and the environment. In equation~\eqref{exe}  the value $X_n$ is a time-dependent function. Then, the Hamiltonian can be represented as
\begin{equation}
\hat H_Q= \sum\limits_n {\omega_{n}^{0} | n \rangle \langle n|}+\sum\limits_n {\alpha_{n}^{'}X_n | n \rangle \langle n|}+
										\sum\limits_{n \ne n'}	{
											\nu_{n,n'} | n \rangle \langle n'|}, 
\label{qua2}
\end{equation}

We will seek the solution of the Schrodinger Eq.~\eqref{shr} in the form
\begin{align}\label{psi}
| \psi \rangle = \sum \limits_{ n} A_{n}(t) | n \rangle .
\end{align}
In expression~\eqref{psi}, the coefficient $A_{n}(t)$,  depends on time, while its squared module determines the probability of the excitation location at the $t$ moment on an $n$-tryptophan.

Substituting the expressions~\eqref{qua2},~\eqref{psi} into the Schrodinger equation~\eqref{shr}, we obtain an expression for the coefficients $A_{n}(t)$. 
\begin{equation}
i \hbar \frac{d A_n}{d t}=\omega_{n}^{0} A_n + \alpha_{n}^{'}X_n A_n + \sum\limits_{n'}	{
											\nu_{n,n'} A_{n'}}.
\label{eqcf}
\end{equation}
Taking into account the interaction only between neighboring tryptophans, we rewrite the expression ~\eqref{eqcf} in the form
\begin{equation}
i \hbar \frac{d A_n}{d t}=\omega_{n}^{0} A_n + \alpha_{n}^{'}X_n A_n +	
											\nu_{n,n-1} A_{n-1} +	
											\nu_{n,n+1} A_{n+1}.
\label{eqcf2}
\end{equation}

Further transformation of the equation~\eqref{eqcf2} will be associated with the introduction of dimensionless quantities $\mu_{n}$, $\mu_{n,n'}$, $W_{m}$, $\kappa_{n}$, $Q_n$, $\tilde{t}$, the values $\rho_n$ and also time parameter $\tau=10^{-14}$ s, where $\tilde{t}=\frac{t}{\tau}$. This transformation is due to the fact that in this case, the system of differential equations obtained by this way is solved numerically much faster with a higher degree of accuracy.
These values are related with the values of the equation~\eqref{eqcf2} by the relations~\cite{lakh} -- \cite{dnaex3},~\cite{41}:
\begin{eqnarray}\label{k}
\omega_{n}^{0} =\frac{\hbar}{\tau}\mu_{n};~~\nu_{n,n'}=\frac{\hbar}{\tau}\mu_{n,n'};~~\kappa_{n}W_{n}^{2}=\frac{\tau^{3}( \alpha_{n}^{'})^{2}}{M \hbar};~~ Q_n=\frac{X_n}{\rho_n};~~ \alpha_{n}^{'}=\frac{\rho_n M}{\tau^2}.
\end{eqnarray}
Then it is easy to show:
\begin{eqnarray}\label{k2}
\alpha_{n}^{'}X_n A_n=\frac{\hbar}{\tau}\kappa_{n}W_{n}^{2}Q_n A_n.
\end{eqnarray}
Using the relations~\eqref{k},~\eqref{k2}, we rewrite the expression~\eqref{eqcf2} as
\begin{equation}
i \frac{d A_n}{d \tilde{t}}=\mu_{n} A_n + \mu_{n,n+1} A_{n+1} +	
											\mu_{n,n-1} A_{n-1} + \kappa_{n}W_{n}^{2}Q_n A_n.	
\label{eqcf3}
\end{equation}
 
As it was defined above, tryptophans in tubulins represent a system of harmonic oscillators obeying Hamilton's equations. We also assume that the movement of this system is damped by the external environment: the molecules that make up the tabulin and the environment external to the tabulins for example the cytosol liquid in which the microtubule is placed inside the cell. In this case, we write the Hamilton equation as
\begin{equation}
M\frac{d^2 X_n}{d t^2}= - \frac{\partial H}{\partial X_n} - \gamma \frac{d X_n}{d t}.
\label{eqg}
\end{equation}
In the expression~\eqref{eqg} $H$ is the complete Hamiltonian of the system: excitation -- tabulins -- external environment; $\gamma$ -- damping coefficient. In the model, the complete Hamiltonian of the system is represented as:
\begin{equation}
H=H_{Cl} +  \langle  \psi|\hat H_Q| \psi \rangle.
\label{gs}
\end{equation}
The second term in the expression~\eqref{gs} determines the average value of the quasiparticle energy in the state $|\psi\rangle$. Using the equalities~\eqref{qua2},~\eqref{psi}, we obtain an expression for the average value
\begin{equation}
\langle  \psi|\hat H_Q| \psi \rangle = \sum\limits_n |A_n|^2\omega_{n}^{0} +   \sum\limits_n |A_n|^2 \alpha_{n}^{'}X_n +  \sum\limits_n (\mu_{n,n+1}A_n^{\star}A_{n+1} + \mu_{n,n-1}A_n^{\star}A_{n-1}).
\label{av}
\end{equation}
Then, using~\eqref{eqg},~\eqref{gs} we get
\begin{equation}
M\frac{d^2 X_n}{d t^2}= -kX_{n} -  \alpha_{n}^{'}|A_n|^2 -  \gamma \frac{d X_n}{d t}.
\label{eqg2}
\end{equation}
Taking into account~\eqref{k} from the expression ~\eqref{eqg2} we will have
\begin{equation}
\frac{d^2 Q_n}{d  \tilde{t}^2}=  -  \gamma^{'} \frac{d Q_n}{d  \tilde{t}}  - W_{n}^{2}Q_n - |A_n|^2,
\label{eqg3}
\end{equation}
where the dimensionless quantities  $\gamma^{'}$ and $W_{n}^{2}$ have the form: $\gamma^{'}=\tau\gamma/M$, $W_{n}^{2}=\tau^2k/M$.

Let 's write down the results obtained in the form of differential equations system
\begin{eqnarray}\label{se}
&&i \frac{d A_n}{d \tilde{t}}=\mu_{n} A_n + \mu_{n,n+1} A_{n+1} +	
											\mu_{n,n-1} A_{n-1} + \kappa_{n}W_{n}^{2}Q_n A_n, \nonumber\\ 
&&\frac{d^2 Q_n}{d  \tilde{t}^2}=  -  \gamma^{'} \frac{d Q_n}{d  \tilde{t}}  - W_{n}^{2}Q_n - |A_n|^2.
\end{eqnarray}

Consider the system~\eqref{se} in more detail. The first three terms in the first equation of the system determine the reversible dynamics of excitation. The term $\kappa_{n}W_{n}^{2}Q_n A_n$ defines the relationship of the excitation dynamics with the classical dynamics of the tryptophans and environment. The mutual influence of classical and quantum dynamics in the model is also determined by the second equation of the system. Thus, in the model, the classical dynamics of the tryptophans and environment affects the quantum dynamics of excitation and vice versa, the quantum dynamics of excitation affects the classical dynamics of tryptophans.

As follows from the equations ~\eqref{se}, the mutual influence of classical and quantum dynamics has an nonlinear nature, thus determining the nonlinear dynamics of the energy transfer mechanisms in microtubules.

\section{Model parameters, initial conditions}
 
The work investigates the migration of the energy along twenty-four tryptophans (Trp1 -- Trp24)  ({\bf Fig.~2}).
It is assumed that  tryptophans displacements can be associated with both the movement of tryptophans inside tabulins, due to their elastic properties, and the movement of tabulins themselves in microtubules. In this regard, the calculations were carried out with the following possible choice of elastic constant $k=5.5$~N/m, $k=4.5$~N/m, $k=1.5$~N/m~\cite{Portet};
the tryptophan mass  $M = 3.4\times 10^{-25}$~kg; damping coefficient  was selected equal to $\gamma=5.3\times 10^{-11}$~kg/s~\cite{ShSu},~\cite{ShSu2}. In this case, the role of a viscous medium, in the work, is performed by cytosol, a liquid, in which the microtubule is suspended inside the cell. 

Interaction with the external environment is determined by choosing of the dimensionless value $\kappa_{n}$~\eqref{k} the same for all tryptophans. In the work, calculations were carried out for the following values $\kappa_{n}$: $0.7$, $0.2$, $0.1$, $0.04$ -- the larger  value of $\kappa_{n}$, the more intense the interaction with the external environment.

It was also assumed that the first tryptophan of the chain was excited at the initial moment of time -- Trp1, which as a result deviated from the equilibrium position by~$0.1$~\AA.

The modeling was carried out for several sets  of tryptophan excitation energies  Table 1 -- Table 3. 

The energy values $\omega_{n}^{0}$ for the first six tryptophans and the values of the dipole-dipole interaction for the tryptophans of each tabulin $\nu_{n,n'}$ were chosen as in work~\cite{get}. Since the energy of the dipole-dipole
interaction between the  tryptophans which are located at the bases of the dimer is unknown, we determined this value as $-50$~sm$^{-1}$ reasoning from the possible geometry of their mutual arrangement.

The precision control was exercised to suit the requirement $|\sum\limits_n |A_n|^2-1| < 1.0\times10^{-5} - |\sum\limits_n |A_n|^2-1| < 1.0\times10^{-2}$. 
\section{Results and discussion}

The probability $|A_n|^2$ of finding the excitation on the $n$ tryptophan depending on the time $t$ is  calculated  in the work.

At the beginning, let's consider the calculation carried out for the sequences of six and eight tryptophans with excitation energies defined in the work~\cite{get}  Table 1.
The calculations are shown in {\bf Fig.~3}~(a), (b) at the different time scales. Here, at the initial moment of time, the excitation with a probability of $1$ was localized on the first tryptophan -- Trp1. Then, $1\times 10^{-11}$~s later, the excitation migrates to the second tryptophan -- Trp2 and localizes on the first two tryptophans within a short period of time. It is obvious from {\bf Fig.~3}~(a) that $2\times 10^{-11}$~s later, the probability of the
excitation location on the first triptophan falls considerably whereas  its location on the  tryptophan Trp2 rises. Then, the probability of excitation presence on the first and second tryptophans fall almost to zero whereas the probability of the excitation existence on the third and fourth tryptophans increases significantly, peaking around $3.5\times 10^{-11}$~s. In the region of $1.5\times 10^{-10}$~s, the probability of excitation staying on the third and fourth tryptophan decreases significantly - excitation is localized on the  fifth and sixth tryptophans. Then the excitation on the fifth tryptophan decreases and reaches its maximum on the sixth tryptophan. Note that the process of excitation migration is an irreversible.  Thus, the signal migrating along the chain reaches the sixth tryptophan in $1.5\times 10^{-10}$~s and localizes on it. The probability of the reverse process is zero. Further migration of excitation is improbable. This is due to the fact that the excitation energy of the seventh tryptophan is greater than that of the sixth, so that the seventh tryptophan represents a potential barrier for the further signal migration. 
In the work~\cite{get}, the possible mechanism of the energy transfer in microtubules was investigated within one tubulin. It was shown starting from the highest energy state, the excitation travels from Trp1 to Trp6  covering the length of the tubulin dimer. 

It is known the biomolecular medium can contribute to the formation of favorable conditions for the transfer of energy through biosystems by influencing for example the potential profile of the latter.
Thus, in the work~\cite{erty} the coupling between the protein environment and the excitation energies of
individual bacteriochlorophyll molecules was determined. In the work~\cite{hjk} the effect of water on the structure and fluorescence spectra of L-tryptophan was investigated. In this regard, see also the work~\cite{uio}.
In that context, the dependence of ionization energies of nitrogenous bases of DNA on the influence of external factors is discussed in the works~\cite{40},~\cite{41},~\cite{kub},~\cite{kub2}.
Thus, Kubar and co-authors~\cite{kub},~\cite{kub2} determined several factors responsible for the charge transfer in DNA: the electrostatic interaction of the charge with  solvent, the
fluctuation of counterions; the DNA base fluctuation leading to the oscillations of ionization potentials in the order of 0.4 eV which has a crucial impact on the energy profile of the DNA's strands.
It seems that in our case, the prevailing factor responsible for the energy transfer in microtubules can be the external electrostatic field (for example, the electrostatic field of microtubules) and the environment in which the microtubules are placed. The latter is cytosol, which can has an appropriate ionic composition and is under the influence of an electrostatic field. Influencing the ionic composition of the microtubule environment, external electrostatic field can play the role of an ordering factor that creates conditions and determines the direction of excitation migration along the tryptophan chain. In the work the conditions under which the energy migration along the microtubule tryptophan chain will be possible are determined.
The corresponding sets of excitation energies are defined in Table 2, Table 3. 

In Table 2, the energy values for tryptophans Trp7 -- Trp12 were selected in such a way that they differed from the excitation energies of tryptophans Trp1 -- Trp6~\cite{get} by no more than $0.056$~eV. Such difference seems quite possible  because it is in accordance with the possible errors in the calculations presented for example in the work ~\cite{ls}, where excitation energies were calculated. 
The results of calculation are presented in {\bf Fig.~4} (a), (b). 
Comparing the results of {\bf Fig.~3} and {\bf Fig.~4}, we see that the process of excitation migration in the latter case is more delayed in time. This is a longer process. Here the interaction with the external environment is less intense:  $\kappa_{n}=0.1$. As before, it is assumed that at the initial moment of time, with a probability of $1$, the excitation was on the first tryptophan. Then the excitation migrates along the chain under consideration, sequentially staying at each group of tryptophans consisting of two or three molecules: Trp1 -- Trp2,  Trp3 -- Trp4,  Trp5 -- Trp6,  Trp6 -- Trp8,  Trp9 -- Trp10,  Trp11 -- Trp12. The excitation reaches the sixth tryptophan in a time approximately equal to
$2.4\times 10^{-9}$~s. Then the excitation passes to the seventh tabulin and migrates along the chain, reaching the twelfth tryptophan in a time of $1.4\times 10^{-8}$~s. It can be seen from {\bf Fig.~3} and {\bf Fig.~4} that the mechanism of excitation migration is of a mixed nature. Thus, within of two or three tryptophans with a strong dipole-dipole interaction, the energy migration process occurs due to an exciton mechanism. So, exciton in the case of {\bf Fig.~4}  is localized on the groups of tryptophans as it was shown above. In this case the excitation that hits the donor molecule passes to neighboring molecules before the relaxation of donor molecule state has time to occur.  In other cases of a weaker dipole-dipole interaction, the donor molecules pass into the ground state, and the acceptor molecule passes into the excited state due to the inductive -- resonant process. These are transitions: Trp2 -- Trp3, Trp4 -- Trp5, Trp8 -- Trp9, Trp10 -- Trp11.

The process shown in {\bf Fig.~5} is even more slowed down. In this case, $\kappa_{n}=0.04$. The interaction with the environment is weaker and the migration process takes place over a longer period of time. So the excitation reaches the sixth tryptophan for the time $9.7\times 10^{-9}$~s.

Consider the process of excitation migration along twenty-four tryptophans arranged in the chain of six tryptophans in each of the four tubulins under consideration {\bf Fig.~6}. The excitation energies for this case are defined in Table 3. Here, the energy values for tryptophans Trp7 -- Trp12 were selected in such a way that they differed from the excitation energy of tryptophans Trp1 -- Trp6 by no more than $0.056~eV$. The energy values for tryptophans Trp13 -- Trp24 were selected so that they differed from the excitation energies of tryptophans Trp1 -- Trp6 by no more than $0.10~-~0.16$~eV.
As before, such a difference quite possible due to the influence of environmental factors.
Here as before it is assumed that at the initial moment of time, the excitation was localized on the first tryptophan with a probability of $1$. Then, during $1.0\times 10^{-9}$~s, the excitation migrates to the fifth and sixth tryptophans. After that, there is a collective excitation  on the sixth, seventh and eight tryptophans. Next, the excitation migrates to the ninth and tenth tryptophan. Then, migration occurs to the eleventh, twelfth, thirteenth and fourteenth molecules. Further, migrating  excitation for short periods of time is localized on the fifteenth and sixteenth tryptophan, after which the seventeenth, eighteenth, nineteenth and twentieth tryptophans experience local excitation. From the twentieth tryptophan, the excitation migrates to the twenty-first~--~twenty-second tryptophans and then to the last couple of tryptophans. Thus, the excitation migrates from the first tryptophan to the twenty-fourth tryptophan within $1.9\times 10^{-8}$~s and localized on the last tryptophan at subsequent moments in time. As can be seen the energy transfer mechanism in this case is of a mixed nature as well.

In conclusion, consider the dynamics of tryptophan excitation energy $\omega_{n}^{0}$  ({\bf Fig.~7}) for the case {\bf Fig.~4} (b). Let us confine ourselves to the consideration of eight tryptophans.Trp1 - Trp8. A comparison between
{\bf Fig.~4} (b) and {\bf Fig.~7}  shows that the energy migration is highly probable on the tryptophan whose excitation energy has a minimum. 
Thus, the probability of finding an excitation on a particular tryptophan migrates following a change in the energy profile of the tryptophan's chain. The latter in turn correlates strongly with the dynamics of 
tryptophan displacement (according to the expression~\eqref{exe}),  in this way demonstrating the dependence of the excitation dynamics on the environment state.

\section{Conclusion}

In conclusion, we note the following. The specificity of the energy transfer mechanisms in microtubules is determined by the nature of tryptophans interaction, as well as the influence of the external environment.

The results obtained in the work allow us to conclude that the mechanism of the energy migration in microtubules can have a mixed character: excitonic, coherent~\cite{get},~\cite{eng}~--~\cite{coli} and inductive-resonant. 

We also note that  the energy migration in microtubules may depend on a number of factors. Thus, the mechanism of the energy migration may depend on the geometry of the microtubule. It is known that a microtubule can be represented as an object consisting of a system of connected tubulins forming a beveled hexagonal lattice having rotational spiral symmetry and possessing anisotropy. As shown, for example, in the works ~\cite{ShSu}, ~\cite{ShSu2}, the specificity of the propagation of mechanical vibrations in a microtubule depends on its geometric and physical inhomogeneities. The dipole-dipole interaction between tryptophans depends on the environment. Besides, the dipole-dipole interaction will be different between tryptophans located on different protophiloments and will also depend on the pathway considered in the microtubule.

Note also that conclusions about the nature of the energy transformation in the tryptophan system may depend on more precise values of the parameters of the model such as the excitation energy $\omega_{0}$, and,  the dipole-dipole interaction $\nu_{n,n'}$ between tryptophans located at the ends of neighboring tubulins. 

An increase of the microtubule geometric dimensions  may also be a significant factor.

As it was noted above, the process of the energy migration in the microtubule can be carried out only after the formation of a the favorable energy profile, namely, the excitation energy of the subsequent tryptophan $\omega_{n}$ should be less than the excitation energy of the previous $n-1$ tryptophan: $\omega_{1} > \omega_{2} > ...  >\omega_{n-1} > \omega_{n} > ...  >\omega_{m}$.
These conditions can be formed either artificially or naturally in the environment surrounding tryptophans in microtubules. This topic deserves a separate detailed discussion, which is beyond the scope of this article.

The results of the work also allow us to talk about a signal function of tryptophans. Therefore the microtubules can be considered as signaling system for transmitting information. Moreover, the energy transfer in microtubules is a quantum phenomenon ~\cite{ham},~\cite{SS21},~\cite{hamer}. The issue becomes particularly relevant in connection with the discussion about the possibility of signal transmission in microtubules without dissipation~\cite{SS21},~\cite{shl}. 

The investigation demonstrates the utilization of the formalism in practical problems for the study of a excitation migration through different molecular sequences that may be tested in experiments.

\section*{Acknowledgements}
We are grateful to A. N. Bugay, O.V. Belov - JINR, for the support of our work. Authors express special gratitude  to Irina Bazhenova for her work on the drawings.


\vspace*{4mm}
\begin{center} 
{\footnotesize{\bf Table 1.} Simulation parameters.\\
\vspace{2mm}
\begin{tabular}{cc} 
 \hline
Tryptophan in the chain& Excitation energy $\omega_{n}^{0}$ in eV\\  
\hline
Trp1&  4.4981   \\

Trp2&  4.4924      \\

Trp3&   4.4808  \\

Trp4&  4.4757    \\

Trp5&  4.4627     \\

Trp6&   4.4500  \\

Trp7&   4.4981   \\

Trp8&   4.4924   \\
\hline
\end{tabular}}\\
\end{center}

\vspace*{4mm}
\begin{center} 
{\footnotesize{\bf Table 2.} Simulation parameters.\\
\vspace{2mm}
\begin{tabular}{cc} 
 \hline
Tryptophan in the chain&  Excitation energy $\omega_{n}^{0}$ in eV\\  
\hline
Trp1&  4.4981     \\

Trp2&  4.4924       \\

Trp3&   4.4808   \\

Trp4&  4.4757     \\

Trp5&  4.4627       \\

Trp6&   4.4500   \\

Trp7&   4,4422   \\

Trp8&   4,4365   \\

Trp9&  4,4246     \\

Trp10&  4,4196       \\

Trp11&   4,4064   \\

Trp12&  4,3938     \\
\hline
\end{tabular}}\\
\end{center}

\vspace*{4mm}
\begin{center} 
{\footnotesize{\bf Table 3.} Simulation parameters.\\
\vspace{2mm}
\begin{tabular}{cc} 
 \hline
Tryptophan in the chain& Excitation energy $\omega_{n}^{0}$ in eV\\  
\hline
Trp1&  4.4981     \\

Trp2&  4.4924       \\

Trp3&   4.4808   \\

Trp4&  4.4757     \\

Trp5&  4.4627       \\

Trp6&   4.4500   \\

Trp7&   4,4422   \\

Trp8&   4,4365   \\

Trp9&  4,4246     \\

Trp10&  4,4196       \\

Trp11&   4,4064   \\

Trp12&  4,3938     \\

Trp13&   4,3863   \\

Trp14&   4,3806   \\

Trp15&   4,3687   \\

Trp16&  4,3637     \\

Trp17&  4,3505       \\

Trp18&   4,3378   \\

Trp19&  4,3299     \\

Trp20&   4,3243   \\

Trp21&  4,3126     \\

Trp22&  4,3075       \\

Trp23&   4,2945   \\

Trp24&  4,2819     \\
\hline
\end{tabular}}\\
\end{center}

\newpage

\begin{center}
\includegraphics{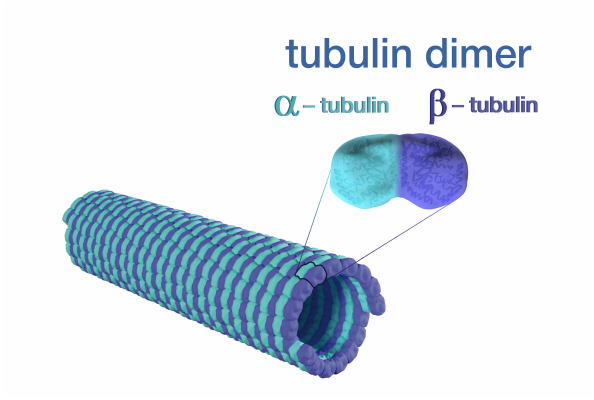}\label{mt.eps}\\[5pt]
\parbox[c]{15cm}{\footnotesize{\bf \hspace{6 cm} Fig.~1.}~Microtubule.}
\end{center}

\begin{center}
\includegraphics{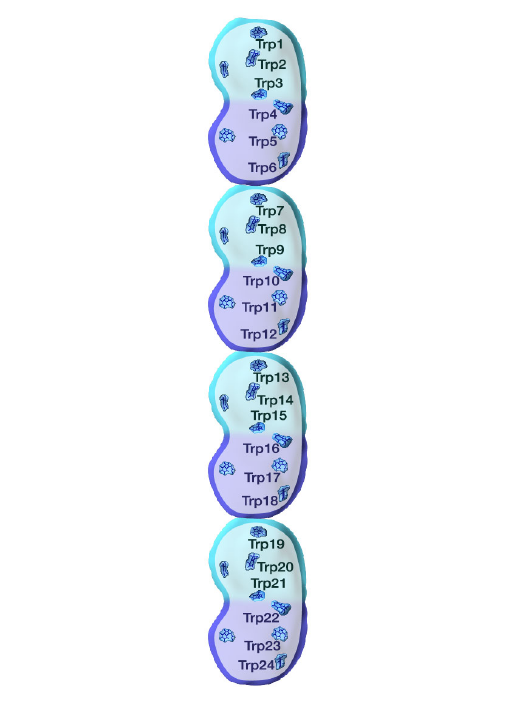}\label{ch}\\[5pt]
\parbox[c]{15cm}{\footnotesize{\bf Fig.~2.}~The set of tryptophans 
in a one-dimension lattice  aligned vertically in the protofilament.}
\end{center}

\begin{center}
\includegraphics{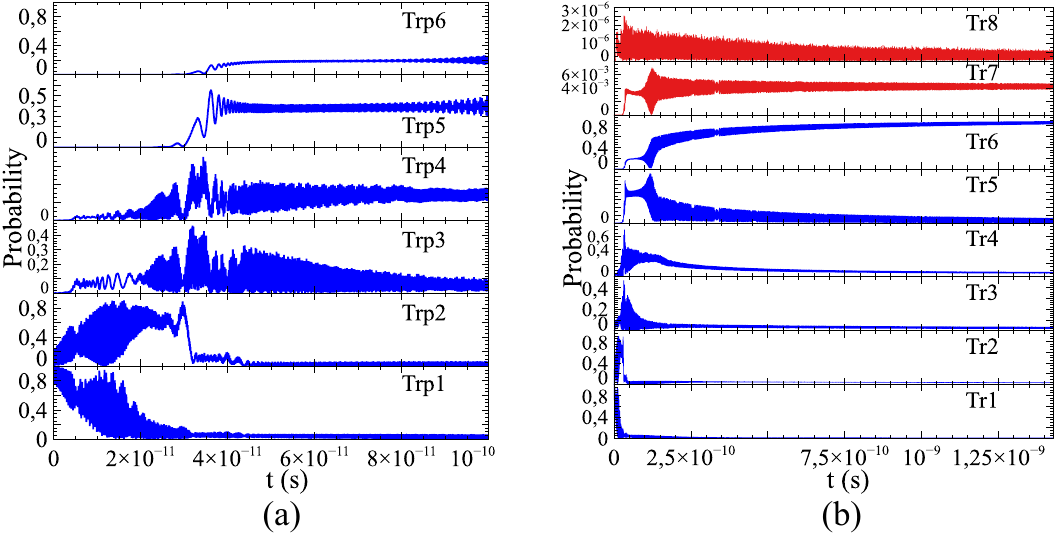}\label{f3ab}\\[5pt]
\hspace*{20mm}\parbox[c]{15cm}{\footnotesize{\bf Fig.~3.}~Time-depending probable excitation location on the tryptophans. $\kappa_{n}=0.7$; $k=5.5~N/m.$}
\end{center}

\begin{center}
\includegraphics{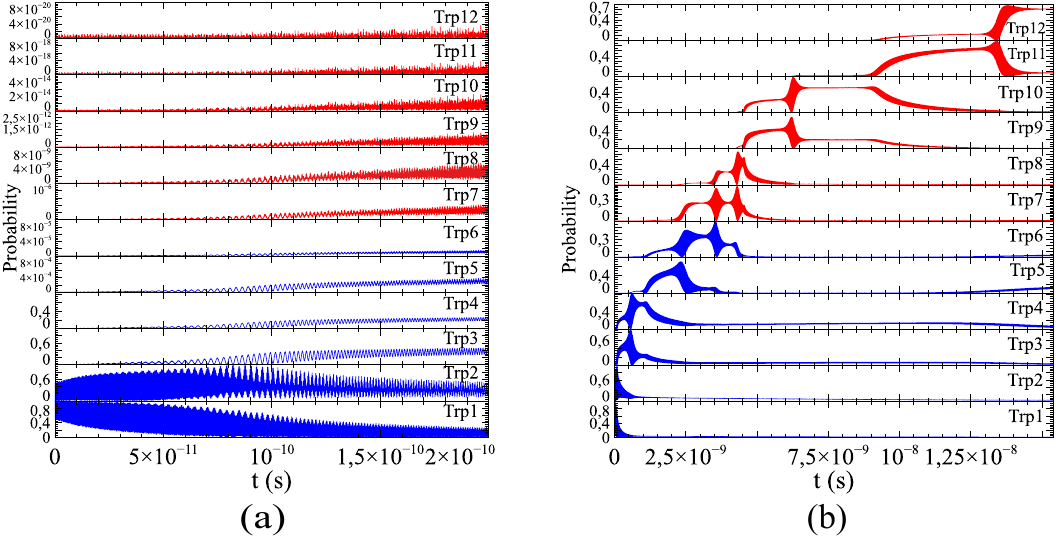}\label{f4ab}\\[5pt]
\hspace*{20mm}\parbox[c]{15cm}{\footnotesize{\bf Fig.~4.}~Time-depending probable excitation location on the tryptophans. $\kappa_{n}=0.1$; $k=4.5~N/m.$}
\end{center}

\begin{center}
\includegraphics{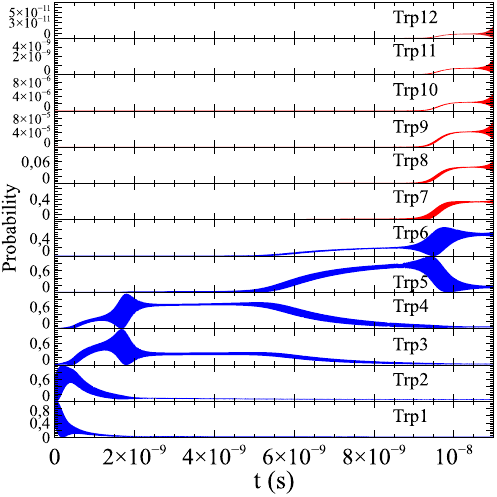}\label{f5}\\[5pt]
\hspace*{17mm}\parbox[c]{15cm}{\footnotesize{\bf Fig.~5.}~Time-depending probable excitation location on 
the tryptophans. $\kappa_{n}=0.04$; $k=4.5~N/m.$}
\end{center}


\begin{center}
\includegraphics{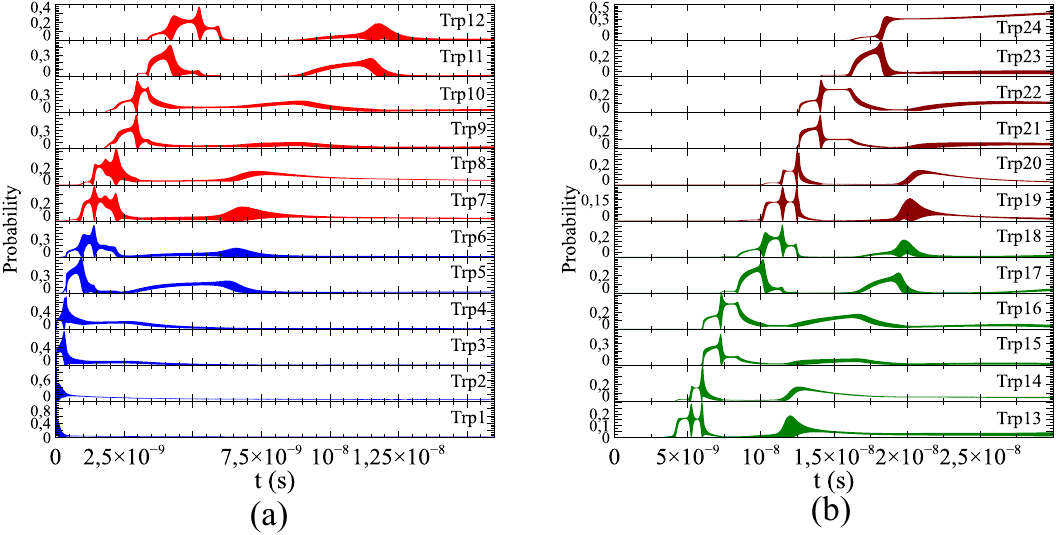}\label{f7}\\[5pt]
\hspace*{20mm}\parbox[c]{15cm}{\footnotesize{\bf Fig.~6.}~Time-depending probable excitation location on the tryptophans. $\kappa_{n}=0.2$; $k=4.5~N/m.$}
\end{center}

\newpage

\begin{center}
\includegraphics{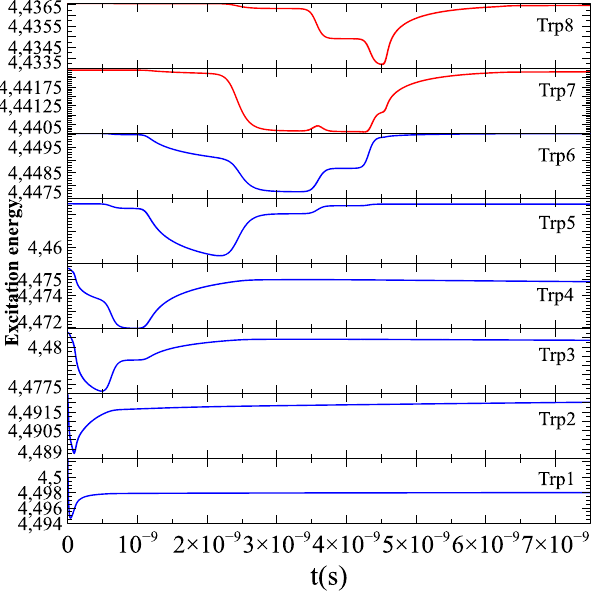}\label{f8}\\[5pt]
\hspace*{52mm}\parbox[c]{20cm}{\footnotesize{\bf Fig.~7.}~Excitation energy (eV). $\kappa_{n}=0.1$; $k=4.5~N/m.$}
\end{center}

\end{CJK*}
\end{document}